\pdfoutput=1

\documentclass[letterpaper,conference]{IEEEtran}

\usepackage{ifpdf}
\usepackage{cite}
\ifCLASSINFOpdf
 	\usepackage[pdftex]{graphicx}
 	\DeclareGraphicsExtensions{.pdf,.jpeg,.png}
\else
 	\usepackage[dvips]{graphicx}
 	\DeclareGraphicsExtensions{.eps}
\fi
\graphicspath{{./figures/}}
\usepackage[cmex10]{amsmath}
\usepackage{amsfonts} 
\usepackage{amssymb} 
\usepackage{amsthm} 
\usepackage{array}
\usepackage[thinlines,thiklines]{easybmat}
\usepackage{url}

\newcommand{\vect}[1]{\mathbf{#1}}
\newcommand{\matr}[1]{\mathbf{#1}}
\newcommand{\set}[1]{\mathcal{#1}}
\newcommand{\code}[1]{\mathcal{#1}}

\newcommand{\Z}{\mathbb{Z}}
\newcommand{\F}{\mathbb{F}}

\newcommand{\stacktwo}[2]{\genfrac{}{}{0pt}{}{#1}{#2}}

\newcommand{\minstar}[1]{\underset{#1}{\operatorname{min}^{*}}}
\newcommand{\wH}{w_{\mathrm{H}}}
\DeclareMathOperator{\wt}{wt}

\DeclareMathOperator{\perm}{perm}

\DeclareMathOperator{\bigO}{O}

\theoremstyle{plain}
\newtheorem{thm}{Theorem} 

\theoremstyle{definition}

\theoremstyle{remark}

\makeatletter
\newlength \figwidth
\if@twocolumn
\else
\fi
\makeatother

\usepackage{hyperref}
\hypersetup{
  pdfinfo={
    Title={Minimum Distances of the QC-LDPC Codes in IEEE 802 Communication Standards},
    Author={Brian K. Butler},
    Subject={Information Theory, Error Correction Coding},
    Keywords={LDPC; Error Correction; Coding; Protograph}
  }
}

\begin{document}
\title{Minimum Distances of the QC-LDPC Codes in IEEE 802 Communication Standards}

\author{
\IEEEauthorblockN{Brian~K.~Butler}%
\IEEEauthorblockA{Butler Research, La~Jolla, CA 92037  butler@ieee.org}%
}

\maketitle

\ifCLASSOPTIONpeerreview
	\markboth{McCoy Theorem for QC-LDPC Codes\quad Version:  November 6, 2014}%
	{Minimum Distance of the QC-LDPC Codes in IEEE 802}
\else
	\markboth{\MakeLowercase{draft for} Conference}
	{Butler: Minimum Distance of the QC-LDPC Codes in IEEE 802}
\fi

%

\begin{abstract}  
This work applies earlier results on Quasi-Cyclic (QC) LDPC codes to the codes specified in six separate IEEE 802 standards, 
specifying wireless communications from $54$ MHz to $60$ GHz.
First, we examine the weight matrices specified to upper bound the codes' minimum distance independent of block length.
Next, we search for the minimum distance achieved for the parity check matrices selected at each block length.
Finally, solutions to the computational challenges encountered are addressed.
\end{abstract}


%
\IEEEpeerreviewmaketitle

\section{Introduction}
\label{sect1}

Since the mid 2000s, LDPC codes have found a wide variety of commercial applications. 
Much about these codes is well understood, but rather frequently little attention is paid to the minimum (Hamming) distance between codewords. 
The minimum distance of a code can limit the error performance at high SNR and is important in understanding the likelihood of undetected errors.
This paper presents the minimum distance of a wide variety of standardized quasi-cyclic (QC) LDPC codes.

LDPC codes are linear block codes, characterized by a sparse parity-check matrix $\matr{H}$.
The set of codewords $\code{C}$ of such a code is defined by the null-space of $\matr{H}$, that is
$\code{C}=\{ \vect{c} \in \F^n : \matr{H}\, \vect{c}^T = \vect{0}^T \}$.
The codewords of a block code may be divided into non-overlapping \emph{subblocks} of $N$ consecutive symbols.
A QC code is a linear block code having the property that applying identical circular shifts to every subblock of a codeword yields a codeword.


\section{Quasi-Cyclic LDPC}
MacKay and Davey introduced upper bounds on the minimum distance for a class of codes that included QC-LDPC codes in \cite{MKseagate}.
Notable later work appeared in \cite{Tan04,FossQC}.
Of particular relevance to this work are the upper bounds of Smarandache and Vontobel in \cite{SmaVon12}.

A binary QC-LDPC code of block length $n = L N$ can be described by an $m \times n$ sparse parity-check matrix $\matr{H}\in \F_2^{m\times n}$, with $m = J N$,
which is composed of $N \times N$ circulant submatrices.
A right \emph{circulant matrix} is a square matrix with each successive row right-shifted circularly one position relative to the row above.
Therefore, circulant matrices can be completely described by a single row or column. 
As in \cite{SmaVon12,ButlerQC}, we use the description corresponding to the left-most column.

In earlier works it was recognized that the set of all $N \times N$ circulant binary matrices form a commutative ring which is isomorphic to the commutative ring of polynomials with binary coefficients modulo $x^N -1$, \textit{i.e.}, $\F_2[x]/\langle{x^N \! - \! 1} \rangle$.
Specification and analysis of the QC-LDPC code can then be carried out on the much smaller polynomial parity-check matrix $\matr{H}(x)$ with entries from the ring,
as described below.

The isomorphism between $N\times N$ binary circulant matrices and polynomial residues in the quotient ring
maps a matrix to the polynomial in which the coefficients in order of increasing degree correspond to 
the entries in the left-most matrix column taken from top to bottom.
Under this isomorphism, the $N\times N$ identity matrix maps to the multiplicative identity in the polynomial quotient ring, namely $1$. A few examples of the mapping (indicated by $\mapsto$) for $N = 3$ are shown below:

\begin{equation*}
\begin{bmatrix}
1\ 0\ 0\\
0\ 1\ 0\\
0\ 0\ 1\end{bmatrix} \mapsto 1 \quad
\begin{bmatrix}
0\ 0\ 1\\
1\ 0\ 0\\
0\ 1\ 0\end{bmatrix} \mapsto x \quad
\begin{bmatrix}
1\ 1\ 0\\
0\ 1\ 1\\
1\ 0\ 1\end{bmatrix} \mapsto 1+x^2.
\end{equation*}


Given a polynomial residue $a(x) \in \F_2[x]/\langle{x^N \!- \!1} \rangle$, we define its weight $\wt(a(x))\in\Z$ to be the number of nonzero coefficients.
Thus, the weight $\wt(a(x))$ of the polynomial $a(x)$ is equal to the Hamming weight $\wH(\vect{a})$ of the corresponding binary vector of coefficients $\vect{a}$.
For a length-$L$ vector of elements in the ring, $\vect{a}(x)=(a_0(x),a_1(x),\ldots,a_{L-1}(x))$, we define its Hamming weight to be the sum of the weights of its components, \textit{i.e.}, $\wH(\vect{a}(x))=\sum_{i=0}^{L-1} \wt(a_i(x))$.
Throughout this work, computations implicitly shift to integer arithmetic upon taking the weight.

The parity-check matrix $\matr{H}$ of a binary QC-LDPC code may be presented in $J \times L$ block matrix form, such that
\begin{equation*}
  {\matr{H}} \triangleq \left[ {\begin{array}{ccc}
     {{\matr{H}}_{0,0} } & \cdots & {{\matr{H}}_{0,L-1} }\\
      \vdots & \ddots & \vdots \\
     {{\matr{H}}_{J-1,0} } & \cdots & {{\matr{H}}_{J-1,L-1} }
   \end{array} } \right],
\end{equation*}
where each submatrix $\matr{H}_{j,i}$ is an $N \times N$ binary circulant matrix.
Let $h_{j,i,s} \in \F_2$ be the left-most entry in the $s$th row of the submatrix $\matr{H}_{j,i}$.
We can then write $\matr{H}_{j,i} = \sum_{s = 0}^{N-1} h_{j,i,s} \matr{I}_s$,
where $\matr{I}_s$ is the $N \times N$ identity matrix circularly left-shifted by $s$ positions.
Now, using the same convention as above for identifying matrices with polynomial residues, we can associate with
$\matr{H}$ the \emph{polynomial parity-check matrix} ${\matr{H}}(x)$, where
${\matr{H}}(x) \in \left(\F_2[x]/\langle{x^N \!-\! 1} \rangle \right)^{J \times L}$,
\begin{equation}
\label{Hx}
 {\matr{H}}(x) \triangleq \left[ {\begin{array}{ccc}
    {h_{0,0} (x)} & \cdots & {h_{0,L-1} (x)}\\
     \vdots & \ddots & \vdots \\
    {h_{J-1,0} (x)} & \cdots & {h_{J-1,L-1} (x)}
  \end{array} } \right],
\end{equation}
and $h_{j,i} (x) \triangleq \sum_{s = 0}^{N-1} {h_{j,i,s} } x^s$.

We will be interested in the weight of each polynomial entry of $\matr{H}(x)$, or, equivalently, the row or column sum of each submatrix of $\matr{H}$.
The \emph{weight matrix} of $\matr{H}(x)$, which is a $J\times L$ matrix of nonnegative integers, is defined as
\begin{equation*}
\wt \left( {{\matr{H}}(x)} \right) \triangleq \left[ {\begin{array}{ccc} 
   {\wt \left( {h_{0,0} (x)} \right)} & \cdots & {\wt \left( {h_{0,L- 1} (x)} \right)} \\
    \vdots & \ddots & \vdots \\
   {\wt \left( {h_{J- 1,0} (x)} \right)} & \cdots & {\wt \left( {h_{J- 1,L- 1} (x)} \right)} \\
 \end{array} } \right].
\end{equation*}
Note that the weight matrix $\matr{A}\triangleq \wt \left( {{\matr{H}}(x)} \right)$ is also
termed the \emph{protomatrix} in the context of protograph-based constructions.

We will use the shorthand notation $[L]$ to indicate the set of $L$ consecutive integers, $\left\{0,1,2,\ldots,L \!-\! 1 \right\}$. 
We let $\set{S}\setminus i$ denote all the elements of $\set{S}$, excluding the element $i$.
We denote by $\matr{A}_\set{S}$ the submatrix of $\matr{A}$ containing the columns indicated by the index set $\set{S}$.
We let $\minstar{}$ denote the usual $\min{}$ function, but let $\minstar{}$ exclude zero values in the argument.

We now have the background and notation to state the theorem which we use to compute the upper bounds on the minimum distance 
that depend on the weight matrices.  This will be used extensively in the remainder of this paper.
\begin{thm}[Theorem 8 in \cite{SmaVon12}]
\label{thm00}
Let $\code{C}$ be a QC code with polynomial parity-check matrix 
$\matr{H}(x) \in \left(\F_2[x]/\langle{x^N \!-\! 1}\rangle\right)^{J \times L}$.
Then the minimum distance of $\code{C}$ satisfies the upper bound
\begin{equation}
\label{distA01}
d_{\min} (\code{C}) \leq \minstar{\stacktwo{\set{S}\subseteq [L]}{|\set{S}| =J+ 1 }}
  \sum_{i \in \set{S}}
  {{\perm \left( {{\matr{A}}_{\set{S}\setminus i}} \right)}}
\quad\quad(\text{in } \Z)
\end{equation}
\end{thm}
\begin{IEEEproof}
See \cite{SmaVon12} and \cite{ButlerQC} for proofs. 
\end{IEEEproof}

\section{Minimum Distances by Standard}
\label{sectMD}
This section introduces a variety of 802-related standards and computes two upper bounds on the minimum distance of their LDPC codes.
The first upper bound is \eqref{distA01} and is based on just the weight matrix. 
Thus, this bound is independent of the code's block length and the polynomials selected for $\matr{H}(x)$.

The second upper bound presented is generally tighter and depends on the precise $\matr{H}$ selected for the standard. 
For this, we conduct a non-exhaustive search for small stopping sets and codewords,
using Richter's algorithm in \cite{Richter06} (there are others).
We are unable to tighten the bounds of these searches by increasing the parameters $I$ and $T$ beyond $150$ and $15$, respectively,
while using pairs of error impulses.
Of course, we modified the generation of impulse locations to take advantage of QC symmetry.
We denote a code's code rate by $r$, where $r=k/n$ for $k$ information bits per block.

\subsection{802.3an Ethernet}
IEEE 802.3an adds a physical layer for $10$ Gigabit Ethernet over unshielded twisted pair copper cabling (10GBASE-T).
A binary $(n,k)=(2048,1723)$ Reed-Solomon-based LDPC code is mandated. 
Its design is described by Djurdjevic \textit{et al.} in \cite{DjurRS}.
It has minimum distance $14$ \cite{SCH13}, but it is not QC .

\subsection{802.11n Wireless LAN}
IEEE 802.11n\cite{80211n} is an amendment to the previous 802.11a/g standards for wireless local area networks in the $2.4$ and $5$ GHz microwave bands.
This amendment adds a ``high throughput'' physical layer specification which encodes the data fields using either a binary $K=7$ convolutional code 
or a QC-LDPC code. 
Support for the convolutional code is mandatory, while the LDPC code is optional.
Four code rates for each of three block length are specified.
The following three tables, grouped by block length, show both upper bounds.
Table~\ref{table_dnl} shows the largest block length codes ($n=1944$ bits, with $N=81$),
Table~\ref{table_dnm} shows the medium block length codes ($n=1296$ bits),
and Table~\ref{table_dns} shows the smallest ($n=648$ bits). 
For rates $1/2$, $2/3$, $3/4$, and $5/6$, the $\matr{H}(x)$ matrices are $12\times24$, $8\times24$, $6\times24$, and $4\times24$, respectively.

\subsection{802.11ad Wireless LAN at $60$ GHz}
IEEE 802.11ad\cite{80211ad} extends the previous wireless local area network standards into the $60$ GHz (\textit{i.e.},``millimeter'' wavelength) band.
This standard contains a ``Directional multi-gigabit'' physical layer specification utilizing QC-LDPC codes to send control and data.
However, a concatenated pair of Reed-Solomon block codes are also specified for an optional low-power mode.
For rates $1/2$, $5/8$, $3/4$, and $13/16$, the $\matr{H}(x)$ matrices are $8\times16$, $6\times16$, $4\times16$, $3\times16$.
In all cases, $n=672$ bits and $N = 42$.
Table~\ref{table_dad} summarizes the distance bounds.

\begin{table}[!t]
\renewcommand{\arraystretch}{1.25} 
\caption{Minimum Distance of 802.11n LDPC Codes with Large Packets}
\label{table_dnl}
\centering
\begin{tabular}{c||c|c}
\hline
\bfseries Code &      \bfseries U.B. of  Weight Matrix &  \bfseries Search using Parity Check \\
\bfseries Rate $r$ &            (independent of $n$) &    {\bfseries Matrix} ($n=1944$ bits)  \\
\hline\hline
$1/2$ & $33$ & $27$\\
$2/3$ & $21$ & $17$\\
$3/4$ & $17$ & $12$\\
$5/6$ & $14$ & $10$\\
\hline
\end{tabular}
\end{table}

\begin{table}[!t]
\renewcommand{\arraystretch}{1.25} 
\caption{Minimum Distance of 802.11n LDPC Codes with Medium-size Packets}
\label{table_dnm}
\centering
\begin{tabular}{c||c|c}
\hline
\bfseries Code &      \bfseries U.B. of Weight Matrix &  \bfseries Search using Parity Check \\
\bfseries Rate $r$ &          (independent of $n$) &  {\bfseries Matrix} ($n=1296$ bits)  \\
\hline\hline
$1/2$ & $31$ & $23$\\
$2/3$ & $17$ & $14$\\
$3/4$ & $22$ & $10$\\
$5/6$ & $17$ & $9$\\
\hline
\end{tabular}
\end{table}

\begin{table}[!t]
\renewcommand{\arraystretch}{1.25} 
\caption{Minimum Distance of 802.11n LDPC Codes with Small Packets}
\label{table_dns}
\centering
\begin{tabular}{c||c|c}
\hline
\bfseries Code &      \bfseries U.B. of Weight Matrix &  \bfseries Search using Parity Check \\
\bfseries Rate $r$ &           (independent of $n$) &  {\bfseries Matrix} ($n=648$ bits)  \\
\hline\hline
$1/2$ & $31$ & $15$\\
$2/3$ & $27$ & $12$\\
$3/4$ & $14$ & $8$\\
$5/6$ & $19$ & $8$\\
\hline
\end{tabular}
\end{table}

\begin{table}[!t]
\renewcommand{\arraystretch}{1.25} 
\caption{Minimum Distance of 802.11ad LDPC Codes}
\label{table_dad}
\centering
\begin{tabular}{c||c|c}
\hline
\bfseries Code &      \bfseries U.B. of Weight Matrix &  \bfseries Search using Parity Check \\
\bfseries Rate $r$ &         (independent of $n$) &  {\bfseries Matrix} ($n=672$ bits)  \\
\hline\hline
$1/2$ & $19$ & $17$\\
$5/8$ & $14$ & $12$\\
$3/4$ & $13$ & $9$\\
$13/16$ & $8$ & $6$\\
\hline
\end{tabular}
\end{table}

\subsection{802.16e Mobile WiMAX and 802.22 Cognitive Wireless}
The IEEE standard 802.16e \cite{80216e}, which was marketed as ``WiMAX'', added mobility to the metropolitan area network (MAN) standards.
Although LDPC codes were included, early WiMAX system deployments used only turbo codes.
Later, the IEEE 802.22 \cite{80222} standard adopted the LDPC codes with minor changes.
This standard is for cognitive wireless regional area networks (RAN) that operate in the TV bands ($54$ -- $862$ MHz).
Table~\ref{table_de} is organized to accommodate the many block lengths included in these two standards at each rate.
For rates $1/2$, $2/3$, $3/4$, and $5/6$, the $\matr{H}(x)$ matrices are $12\times24$, $8\times24$, $6\times24$, and $4\times24$, respectively.

For the largest block size $n=2304$ bits, the polynomials are directly specified to create each $\matr{H}(x)$.
For the smaller block sizes, the standard uses two techniques to create the polynomials for the reduced-sized rings based on the largest block polynomials: proportional scaling and modulo scaling of each exponent. 
The distances presented for 802.16e match those by Rosnes \textit{et al.}, in their updated paper \cite{RosnesYtrehusAdd}.

\subsection{WiMedia UWB (formerly 802.15.3a)}
IEEE 802.15.3a was formed to provide an ultra wideband (UWB) physical layer offering wireless speeds of 480 Mbits/s over short-ranges.
The task group was officially disbanded due to the inability of parties to reach consensus between the multiband OFDM and direct sequence UWB proposals.
The members of the WiMedia Alliance standardized the multiband OFDM technology using $K=7$ convolutional codes initially.
This technology specifies a signal approximately $500$ MHz in bandwidth hopping across the $3.1$ -- $10.6$ GHz frequency band.

Version 1.5 of the WiMedia UWB specification \cite{wimedia} now also specifies QC-LDPC codes. 
More precisely, LDPC coding is an option for all rates from $160$ to $480$ Mbits/s and is required at rates above $480$ Mbits/s.
Table~\ref{table_duwb} shows our findings for the four rates of this standard: $1/2$, $5/8$, $3/4$, and $4/5$.
The $\matr{H}(x)$ matrices are quite large, at $20\times40$, $15\times40$, $10\times40$, and $8\times40$, respectively.
Their size has significantly limited our ability to complete the computation of \eqref{distA01}.
The four rates mentioned here are known as the ``fundamental'' code rates of the standard.
Each code rate has an ``expanded version'' with additional parity, which adds four rows and columns to $\matr{H}(x)$.  (The expanded versions are not analyzed.)

\subsection{802.15.3c millimeter WPAN}
IEEE 802.15.3c has standardized a ``mmWave'' PHY for wireless personal area networks (WPAN) for $60$ GHz in \cite{802153c}.
It is intended for short-range communication and contains a variety of coding schemes.
Our bounds appear in Table~\ref{table_dc} for the four rates of LDPC codes.
For rates $1/2$, $3/4$, $7/8$, and $14/15$, the $\matr{H}(x)$ matrices are $16\times32$, $8\times32$, $4\times32$, and $1\times15$, respectively.
The final one, for rate $14/15$, is unique within this work, as each entry in \eqref{Hx} is the summation of three cyclic permutation matrices.
The entries of all other $\matr{H}(x)$ matrices referenced herein are cyclic permutation or zero matrices and their corresponding weight 
matrices are of course binary.

\begin{table}[!t]
\renewcommand{\arraystretch}{1.25} 
\caption{Minimum Distance of 802.16e and 802.22 LDPC Codes}
\label{table_de}
\centering
\begin{tabular}{c||c|c|c|c|c|c}
\hline
\bfseries Block Length & \multicolumn{6}{c}{\bfseries Code Rate $r$}\\
\cline{2-7}
          $n$ (in bits)  & $1/2$ &	$2/3$A\rlap{\textsuperscript{a}}&	$2/3$B &	$3/4$A &	$3/4$B\rlap{\textsuperscript{a}} &	$5/6$ \\
\hline\hline
$384\rlap{\textsuperscript{b}}$ &	$15$ &	$8$ &	$8$ &	$8$ &	$7$ &	$4$ \\
$480\rlap{\textsuperscript{b}}$ &	$16$ &	$10$ &	$12$ &	$7$ &	$8$ &	$6$ \\
$576$ &	$13$ &	$10$ &	$12$ &	$10$ &	$8$ &	$5$ \\
$672$ &	$19$ &	$9$ &	$11$ &	$10$ &	$8$ &	$7$ \\
$768$ &	$20$ &	$8$ &	$14$ &	$10$ &	$9$ &	$7$ \\
$864$ &	$19$ &	$11$ &	$16$ &	$12$ &	$11$ &	$7$ \\
$960$ &	$19$ &	$13$ &	$15$ &	$12$ &	$11$ &	$7$ \\
$1056$ &	$21$ &	$10$ &	$15$ &	$13$ &	$9$ &	$7$ \\
$1152$ &	$19$ &	$14$ &	$16$ &	$13$ &	$11$ &	$7$ \\
$1248$ &	$22$ &	$13$ &	$15$ &	$13$ &	$9$ &	$7$ \\
$1344$ &	$23$ &	$14$ &	$16$ &	$14$ &	$12$ &	$7$ \\
$1440$ &	$27$ &	$13$ &	$17$ &	$12$ &	$10$ &	$7$ \\
$1536$ &	$20$ &	$12$ &	$15$ &	$14$ &	$11$ &	$7$ \\
$1632$ &	$27$ &	$13$ &	$18$ &	$13$ &	$13$ &	$7$ \\
$1728$ &	$21$ &	$15$ &	$15$ &	$17$ &	$13$ &	$8$ \\
$1824$ &	$19$ &	$15$ &	$15$ &	$13$ &	$12$ &	$8$ \\
$1920$ &	$25$ &	$15$ &	$16$ &	$17$ &	$10$ &	$7$ \\
$2016$ &	$27$ &	$15$ &	$15$ &	$17$ &	$12$ &	$7$ \\
$2112$ &	$28$ &	$15$ &	$16$ &	$15$ &	$14$ &	$8$ \\
$2208$ &	$23$ &	$15$ &	$20$ &	$18$ &	$13$ &	$8$ \\
$2304$ &	$31$ &	$15$ &	$15$ &	$19$ &	$12$ &	$9$ \\
\hline
\bfseries U.B. of &&&&& \\
\bfseries Weight Matrix&	$32$ &	$23$ &	$45$ &	$28$ &	$20$ &	$14$ \\
\hline
\multicolumn{7}{l}{\scriptsize \textsuperscript{a}Configuration appears in IEEE 802.16e, but not in 802.22.}\\
\multicolumn{7}{l}{\scriptsize \textsuperscript{b}Configuration appears in IEEE 802.22, but not in 802.16e.}
\end{tabular}
\end{table}

\begin{table}[!t]
\renewcommand{\arraystretch}{1.25} 
\caption{Minimum Distance of WiMedia UWB LDPC Codes}
\label{table_duwb}
\centering
\begin{tabular}{c||c|c}
\hline
\bfseries Code &      \bfseries U.B. of Weight Matrix &  \bfseries Search using Parity Check \\
\bfseries Rate $r$ &      (independent of $n$)   &    {\bfseries Matrix} ($n=1200$ bits) \\
\hline\hline
$1/2$ & $43$\rlap{\textsuperscript{c}} & $26$ \\
$5/8$ & $35$\rlap{\textsuperscript{c}} & $17$ \\
$3/4$ & $18$ & $9$ \\
$4/5$ & $18$ & $7$ \\
\hline
\multicolumn{3}{l}{\scriptsize \textsuperscript{c}Due to complexity, matrix analysis is only $0.87\%$ and $22\%$ complete to date.}\\
\end{tabular}
\end{table}

\begin{table}[!t]
\renewcommand{\arraystretch}{1.25} 
\caption{Minimum Distance of 802.15.3c LDPC Codes}
\label{table_dc}
\centering
\begin{tabular}{c||c|c c}
\hline
\bfseries Code &      \bfseries U.B. of Weight Matrix & \multicolumn{2}{c}{ \bfseries Search using Parity  }\\
\bfseries Rate $r$ &      (independent of $n$)   & \multicolumn{2}{c}{\bfseries Check Matrix}  \\
\hline\hline
$1/2$ & $36$ & $18$ & ($n=672$ bits)\\
$3/4$ & $16$ & $10$ & ($n=672$ bits)\\
$7/8$ & $16$ & $6$  & ($n=672$ bits)\\
$14/15$ & $6$ & $6$ & ($n=1440$ bits)\\
\hline
\end{tabular}
\end{table}

\section{Improving Computation Time}
In this work and our earlier work on the AR4JA code in \cite{ButlerQC}, we have found that the computation time required 
for \eqref{distA01} is frequently more than the search time of Richter's algorithm, at the block lengths studied here (\textit{e.g.}, $n<4000$ bits).
This is especially true for larger values of $L$.
This section summarizes our efforts to speed-up the computations.

While quite similar in their definition, the matrix permanent and determinant are very different computational tasks.
It is a problem well studied in combinatorics and computer algorithms, but few ready routines are published online.
Also, our application frequently favors sparse matrices whose elements are often $\{0,1\}$.
Without loss of generality, the focus of our implementation will be on the MATLAB computation environment while using C-programming for the permanent functions in an integrated way known as C-MEX.

\subsection{Algorithms for Computing the Matrix Permanent}
Clearly, computing \eqref{distA01}  involves taking the permanent of many submatrices.
By increasing the size of the permanent's argument, we can reformulate \eqref{distA01} to
\begin{equation}
\label{distA02}
d_{\min} (\code{C}) \leq \minstar{\stacktwo{\set{S}\subseteq [L]}{|\set{S}| =J+ 1 }}
    {\perm \left(
    \begin{bmatrix}  
          {\matr{A}}_{\set{S}} \\
          1 \ldots 1              
    \end{bmatrix} \right)}
\quad\quad(\text{in } \Z).
\end{equation}
The summation within \eqref{distA01} can be viewed as a cofactor expansion of the larger matrix permanent in \eqref{distA02} along the bottom row, which is all-ones\footnote{In the case of punctured codes, such as AR4JA, ones in the bottom row of \eqref{distA02} may be replaced by zeros for those columns which correspond to punctured symbols following the puncturing arguments developed in \cite{ButlerQC}.}.

A simple method to compute the permanent is by cofactor expansion (\textit{i.e.}, Laplace expansion), which recursively computes the permanents of smaller and smaller matrices. This method requires about $N!N/2$ operations for a dense $N\times N$ matrix. 
When the matrix is sparse, as happens often in our application, the recursion can be truncated saving significant time.
Our first work on this subject, in \cite{ButlerQC}, relied on just such an algorithm and followed \eqref{distA01}.
We have made several advancements since then.

The first improvement was simple. Previously, for each recursive call, a new submatrix was created with the appropriate rows and columns.
We realized this was wasteful and that simply keeping track of which rows and columns were removed and keeping the matrix unchanged in memory would be more efficient. 
This realized an overall speed-up by a factor of about three.

The best known efficient algorithm for computing general permanents is by Ryser, in\cite{RyserPerm}. 
It is based on the principles of inclusion-exclusion.
Ryser's method requires about $2^N N^2$ operations using standard ordering and $2^N N$ using an incremental Gray-coded approach.
A further modification to Ryser's algorithm by Nijenhuis and Wilf improves the computations' speed by about a factor of two \cite{Nijenhuis}.

\begin{figure}[!t]
\centering
\includegraphics[width=\figwidth]{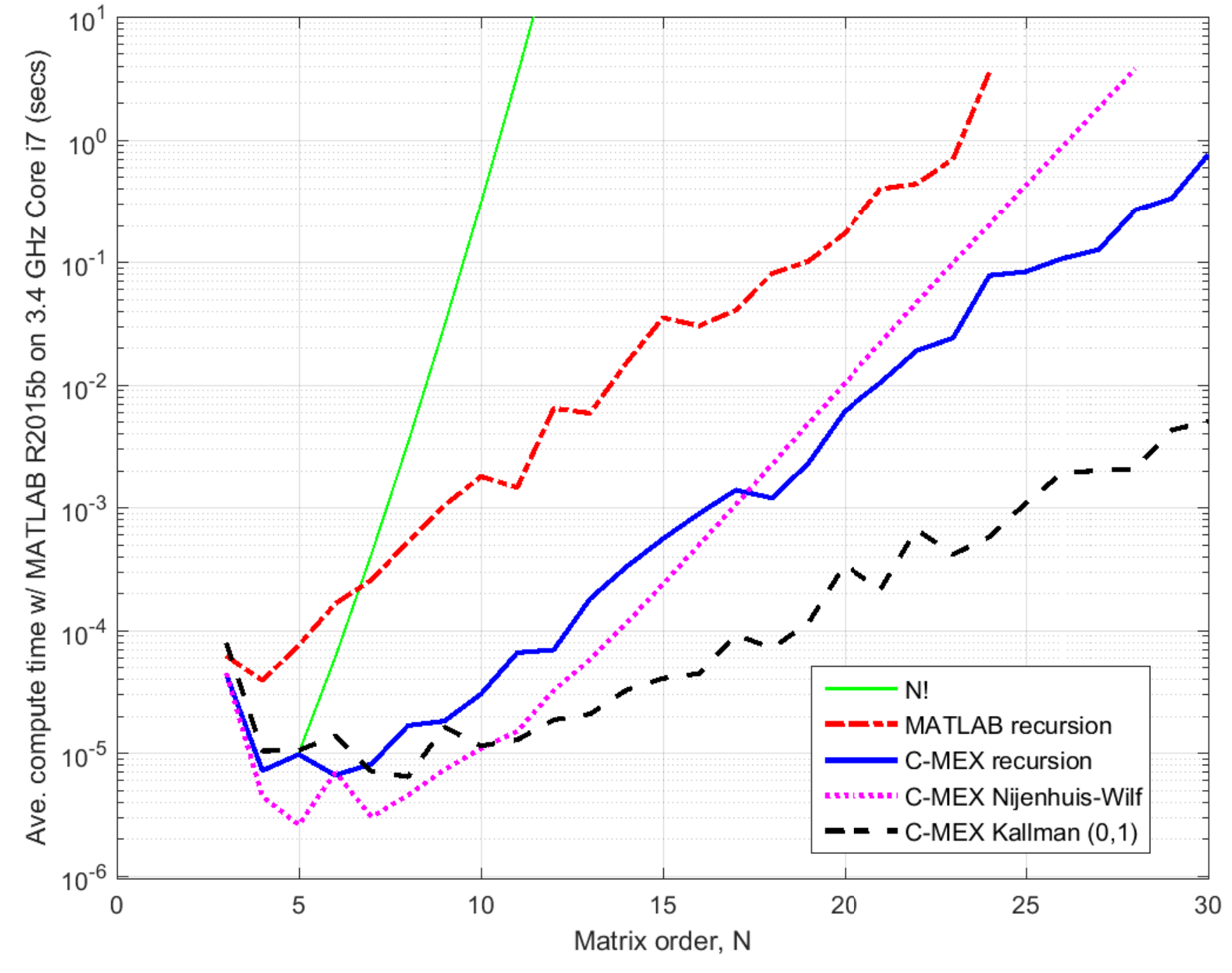}
\caption{Average time (in secs.) to compute the permanent of $5$ random sparse $\{0,1\}$-matrices of column weight $3$ vs.\ matrix order $N$.}
\label{fig_permtime3}
\end{figure}

\begin{figure}[!t]
\centering
\includegraphics[width=\figwidth]{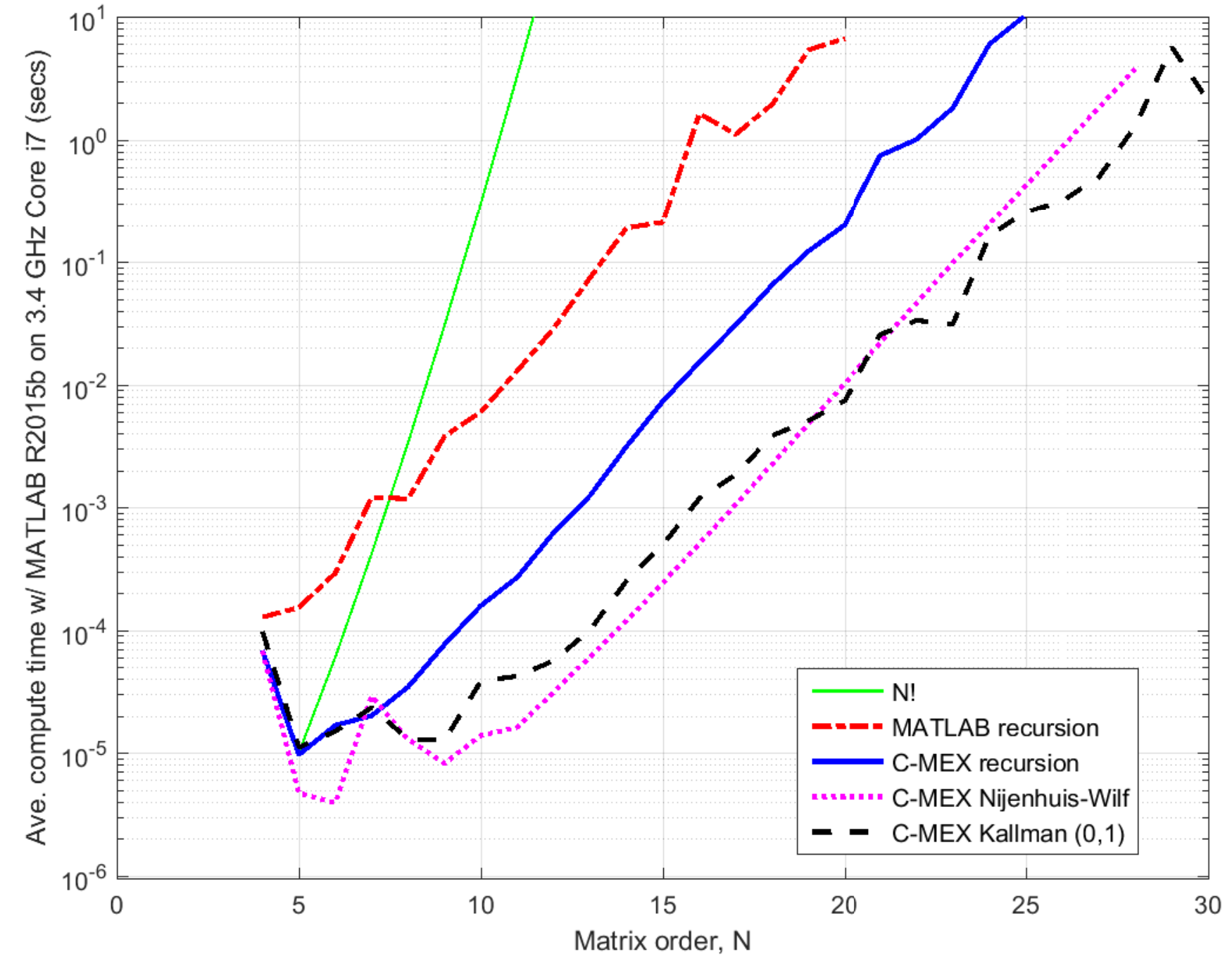}
\caption{Average time (in secs.) to compute the permanent of $5$ random sparse $\{0,1\}$-matrices of column weight $4$ vs.\ matrix order $N$.}
\label{fig_permtime4}
\end{figure}

We have also found a fast permanent algorithm for $\{0,1\}$-matrices by R. Kallman in \cite{Kallman},
which uses row operations and combinatorics to reduce the complexity substantially. 
It is particularly suited for sparse matrices or matrices with certain row relationships.

Since the permanent algorithms are no worse than $\bigO(N!)$ we find it worthwhile to shift from \eqref{distA01} to \eqref{distA02}.
Such a switch has additional gains in less C-MEX calling overhead.
However, making such a switch has the effect of increasing the density of the matrix argument to the permanent function.
As the computation time of our recursive algorithm is heavily dependent upon the density of the matrix, we find that it is typically
no longer advantageous to use it. 

The three main permanent routines mentioned are made available online at \cite{butlerfilex} and compared in Figs.~\ref{fig_permtime3} and \ref{fig_permtime4}.
These results use random sparse matrices where the only constraints are the specified column weight and a non-zero permanent.
While the recursive routine (solid blue) is competitive with Nijenhuis-Wilf (dotted magenta) for column weight $3$ in Fig.~\ref{fig_permtime3}, it is much slower for column weight $4$ in Fig.~\ref{fig_permtime4}.
While the speed of Nijenhuis-Wilf and Kallman (dashed black) were comparable for column weight $4$ as shown in Fig.~\ref{fig_permtime4}, we find that Nijenhuis-Wilf is faster for column weight $5$, not shown.

\subsection{Parallel Processing}
Since MATLAB has supported parallel computing for a number of years, we undertook an effort to incorporate it.
There are several communication constraints between the processes.
The child processes do not communicate with each other and the child processes do not easily return intermediate results to the parent process until they all complete.
Thus, the simplest solution is to make a hierarchy of loops. 

The lowest loop contains sufficient iterations to be a self-contained chunk of processing for a child.
The middle loop utilizes the MATLAB \texttt{parfor} statement which works much like a \texttt{for} loop, but iterations are parallelized.
Due to apparent overhead issues, we have set the number of iterations of this middle loop to be several times the number of hardware processors.
Only upon the completion of the middle loop can intermediate results be aggregated from the children due to the constraints noted above.
Thus, we prefer that the duration of the middle loop's iterations be on the order of $30$ minutes, so that intermediate results may be monitored and saved permanently. 
The highest level loop then runs sufficiently long to exhaust all set combinations of \eqref{distA02}, which may stretch to weeks in some cases.
When running on the four independent processors in our 5th generation Intel Core i7 processor, we typically note a speed increase by a factor of three.
These techniques would scale to even larger parallel computing environments supported by MATLAB.

Iterations proceed through an ${L \choose J+1}$ \textit{combinatorial numbering system}, where each calculation is uniquely identified by the $J+1$ members of the subset $\set{S}$
and the subsets are ordered lexicographically.
To formulate an appropriate starting subset for each processing chunk, we implemented the $\mathrm{unrank}()$ operation which quickly
translates numbering by integers to the correct subset.
The $\mathrm{unrank}()$ and $\mathrm{rank}()$ routines are also made available at \cite{butlerfilex}.

\section{Conclusions}
We presented the minimum distances of a variety of QC-LDPC codes appearing in the IEEE 802-related standards.
As many of the weight matrices analyzed were quite large, we have presented a simplification to the upper bound equation 
and computational optimizations. 
We are now able to compute these distance bounds at least $100$ times faster than just a few years ago.

\section*{Acknowledgment} 

The author would like to thank Michal Kvasnicka, of \'UJV \v{R}e\v{z}, a. s., for permanent references and discussions.



\bibliographystyle{IEEEtran}
\bibliography{IEEEabrv,Butler}

\begin{thebibliography}{10}
\providecommand{\url}[1]{#1}
\csname url@samestyle\endcsname
\providecommand{\newblock}{\relax}
\providecommand{\bibinfo}[2]{#2}
\providecommand{\BIBentrySTDinterwordspacing}{\spaceskip=0pt\relax}
\providecommand{\BIBentryALTinterwordstretchfactor}{4}
\providecommand{\BIBentryALTinterwordspacing}{\spaceskip=\fontdimen2\font plus
\BIBentryALTinterwordstretchfactor\fontdimen3\font minus
  \fontdimen4\font\relax}
\providecommand{\BIBforeignlanguage}[2]{{%
\expandafter\ifx\csname l@#1\endcsname\relax
\typeout{** WARNING: IEEEtran.bst: No hyphenation pattern has been}%
\typeout{** loaded for the language `#1'. Using the pattern for}%
\typeout{** the default language instead.}%
\else
\language=\csname l@#1\endcsname
\fi
#2}}
\providecommand{\BIBdecl}{\relax}
\BIBdecl

\bibitem{MKseagate}
D.~J.~C. MacKay and M.~C. Davey, ``Evaluation of {G}allager codes for short
  block length and high rate applications,'' in \emph{Codes, Systems and
  Graphical Models (Minneapolis, MN, 1999)}, B.~Marcus and J.~Rosenthal,
  Eds.\hskip 1em plus 0.5em minus 0.4em\relax New York: Springer-Verlag, 2000,
  pp. 113--130.

\bibitem{Tan04}
R.~M. Tanner, D.~Sridhara, A.~Sridharan, T.~E. Fuja, and D.~J. Costello, Jr.,
  ``{LDPC} block and convolutional codes based on circulant matrices,''
  \emph{{IEEE} Trans. Inf. Theory}, vol.~50, no.~12, pp. 2966--2984, Dec. 2004.

\bibitem{FossQC}
M.~P.~C. Fossorier, ``Quasi-cyclic low-density parity-check codes from
  circulant permutation matrices,'' \emph{{IEEE} Trans. Inf. Theory}, vol.~50,
  no.~8, pp. 1788--1793, Aug. 2004.

\bibitem{SmaVon12}
R.~Smarandache and P.~O. Vontobel, ``Quasi-cyclic {LDPC} codes: Influence of
  proto- and {T}anner-graph structure on minimum {H}amming distance upper
  bounds,'' \emph{{IEEE} Trans. Inf. Theory}, vol.~58, no.~2, pp. 585--607,
  Feb. 2012.

\bibitem{ButlerQC}
B.~K. Butler and P.~H. Siegel, ``Bounds on the minimum distance of punctured
  quasi-cyclic {LDPC} codes,'' \emph{{IEEE} Trans. Inf. Theory}, vol.~59,
  no.~7, pp. 4584--4597, Jul. 2013.

\bibitem{Richter06}
G.~Richter, ``Finding small stopping sets in the {T}anner graphs of {LDPC}
  codes,'' in \emph{Proc.\ 4th Int. Symp. on Turbo Codes}, Munich, Germany,
  Apr. 2006, pp. 1--5.

\bibitem{DjurRS}
I.~Djurdjevic, J.~Xu, K.~Abdel-Ghaffar, and S.~Lin, ``Class of low-density
  parity-check codes constructed based on {R}eed-{S}olomon codes with two
  information symbols,'' \emph{{IEEE} Commun. Lett.}, vol.~7, no.~7, pp.
  317--319, Jul. 2003.

\bibitem{SCH13}
S.~Zhang and C.~Schlegel, ``Controlling the error floor in {LDPC} decoding,''
  \emph{{IEEE} Trans. Commun.}, vol.~61, no.~9, pp. 3566--3575, Sep. 2013.

\bibitem{80211n}
\emph{{IEEE} Standard for Information Technology-- Local and metropolitan area
  networks-- Specific requirements-- Part 11: Wireless {LAN} Medium Access
  Control ({MAC}) and Physical Layer ({PHY}) Specifications Amendment 5:
  Enhancements for Higher Throughput}, {IEEE} Std. 802.11n-2009, Oct 29, 2009.

\bibitem{80211ad}
\emph{{IEEE} Standard for Information Technology-- Local and metropolitan area
  networks-- Specific requirements-- Part 11: Wireless {LAN} Medium Access
  Control ({MAC}) and Physical Layer ({PHY}) Specifications Amendment 3:
  Enhancements for Very High Throughput in the 60 GHz Band}, {IEEE} Std.
  802.11ad-2012, Dec 28, 2012.

\bibitem{80216e}
\emph{{IEEE} Standard for Local and metropolitan area networks-- Part 16: Air
  Interface for Fixed and Mobile Broadband Wireless Access Systems Amendment 2:
  Physical and Medium Access Control Layers for Combined Fixed and Mobile
  Operation in Licensed Bands and Corrigendum 1}, {IEEE} Std. 802.16e-2005 and
  802.16-2004/Cor 1-2005, Feb 28 2006.

\bibitem{80222}
\emph{{IEEE} Standard for Information Technology-- Local and metropolitan area
  networks-- Specific requirements-- Part 22: Cognitive Wireless {RAN} Medium
  Access Control ({MAC}) and Physical Layer ({PHY}) specifications: Policies
  and procedures for operation in the {TV} Bands}, {IEEE} Std. 802.22-2011,
  July 1 2011.

\bibitem{RosnesYtrehusAdd}
E.~Rosnes, {\O}.~Ytrehus, M.~A. Ambroze, and M.~Tomlinson, ``Addendum to `{An}
  efficient algorithm to find all small-size stopping sets of low-density
  parity-check matrices','' \emph{{IEEE} Trans. Inf. Theory}, vol.~58, no.~1,
  pp. 164--171, Jan. 2012.

\bibitem{wimedia}
\emph{MultiBand {OFDM} Physical Layer Specification}, {WiMedia} Alliance, Inc.
  Std. Final Deliverable 1.5, Aug 11 2009.

\bibitem{802153c}
\emph{{IEEE} Standard for Information Technology-- Local and metropolitan area
  networks-- Specific requirements-- Part 15.3: Amendment 2:
  Millimeter-wave-based Alternative Physical Layer Extension}, {IEEE} Std.
  802.15.3c-2009, Oct 12 2009.

\bibitem{RyserPerm}
H.~J. Ryser, \emph{Combinatorial Mathematics}, ser. The Carus Math.\
  Monographs.\hskip 1em plus 0.5em minus 0.4em\relax The Mathematical
  Association of America, 1963, vol.~14.

\bibitem{Nijenhuis}
A.~Nijenhuis and H.~S. Wilf, \emph{Combinatorial Algorithms for Computers and
  Calculators}.\hskip 1em plus 0.5em minus 0.4em\relax New York, NY: Academic
  press, 1978, ch.~23.

\bibitem{Kallman}
R.~Kallman, ``A method for finding permanents of 0,1 matrices,''
  \emph{Mathematics of Computation}, vol.~38, no. 157, Jan. 1982.

\bibitem{butlerfilex}
\BIBentryALTinterwordspacing
B.~K. Butler. (2015) Files on {MATLAB Central}: File exchange. [Online].
  Available:
  \url{http://www.mathworks.com/matlabcentral/fileexchange/?term=authorid:125678}
\BIBentrySTDinterwordspacing

\end{thebibliography}

\end{document}